\documentclass[11pt,twoside]{article}
\usepackage{asp2014}

\aspSuppressVolSlug
\resetcounters

\begin{document}

\title{Pioneering the Exascale era with Astronomy}

\author{J.~B.~R.~Oonk$^{1}$, C.~Schrijvers$^{1}$, Y.~van~den~Berg$^{1}$}
\affil{$^1$SURF/SURFsara, P.O. Box 94613, 1090 GP Amsterdam, The Netherlands; \email{raymond.oonk@surfsara.nl}}

\paperauthor{J. B. Raymond~Oonk}{}{ORCID}{SURF}{SURFsara}{Amsterdam}{Noord Holland}{1090 GP}{The Netherlands}



  
\begin{abstract}
SURF is the collaborative ICT organisation for Dutch education \& research and coordinates the national e-infrastructure (surf.nl). To accelerate scientific discovery, SURF invests in, operates and explores high-end IT solutions for and with researchers in the Netherlands. In this ADASS 2019 contribution we present our latest developments in high performance and high throughput cloud computing. These developments are particularly relevant for Astronomy, as this science domain: (i) generates large (Petabyte sized) data collections, (ii) uses rapid release and deployment schemes for their software, (iii) requires flexible and interactive test and staging environments and (iv) needs to execute complex workflows on diverse data structures.

We highlight our new OpenStack-based, cloud infrastructure layer for data processing. This allows us to efficiently build, deploy and scale infrastructure as a service (IaaS) \& managed platforms as a service (PaaS) that may be tailored to the specific needs of individual research projects or scientific domains. It also enables dedicated and customizable use of IT resources by different user groups. Currently SURF supports several PaaS solutions for data processing. Examples are, our Grid processing facility, our Kubernetes cluster deployments, and Spider, our new addition to SURF's high throughput data processing (HTDP) platforms. 

Here we focus on Spider, offering HTDP users a rich ecosystem and enabling both batch and interactive data processing at scale. Spider is seamlessly integrated with a large variety of storage systems and tightly coupled with related services via a powerful EVPN network. Modern IT ecosystems, such as Spider, that are created from re-usable building blocks are particularly relevant for the current and next generation of data-intensive astronomical experiments and the diverse need of their users. These experiments include, but are not limited to, the Square Kilometre Array (SKA), the Low Frequency Array (LOFAR) and the Atacame Large Millimetre Array (ALMA).
\end{abstract}

\section{The SURF Virtual Data Center}\label{VDC}
In the last decades the amount and complexity of data generated and analysed has grown exponentially and this is expected to continue. In the next decade (2020--2030) individual instruments, such as the High Luminosity Large Hadron Collider and the SKA, will generate scientific data products in excess of 1~Exabyte per year. Their precursors already operate at the Petabyte-scale per year and have developed their own IT solutions that often require very specific platforms on dedicated hardware to operate, maintain and use. Such a ''pillar'' approach towards IT for research is no longer feasible, nor is it efficient on the scales that current and future experiments operate at. 

To increase this efficiency we need to optimise the usage of the available IT resources, for example, by sharing them amongst research projects and domains. At the same time there is no single, common IT platform that will easily fit all use-cases. Different projects will experience different pains in twisting their applications, via a single narrow platform spanning layer to the underlying hardware pool. This rigidity in interfacing user applications with the hardware in a shared environment is, for example, well described by the hourglass model in \citet{beck2016hourglass}.  

At SURF we recognise this rigidity and aim to resolve (some of) it. For data processing applications we have created the SURF Virtual Data Center (VDC) using cloud technology (Fig.~\ref{fig1}). In 2019 we use OpenStack (https://www.openstack.org) with Terraform (https://www.terraform.io) to manage our VDC cloud resources and provide these as PaaS solutions to our research communities. In the future we will extend this with IaaS solutions and GPU access.

\articlefigure{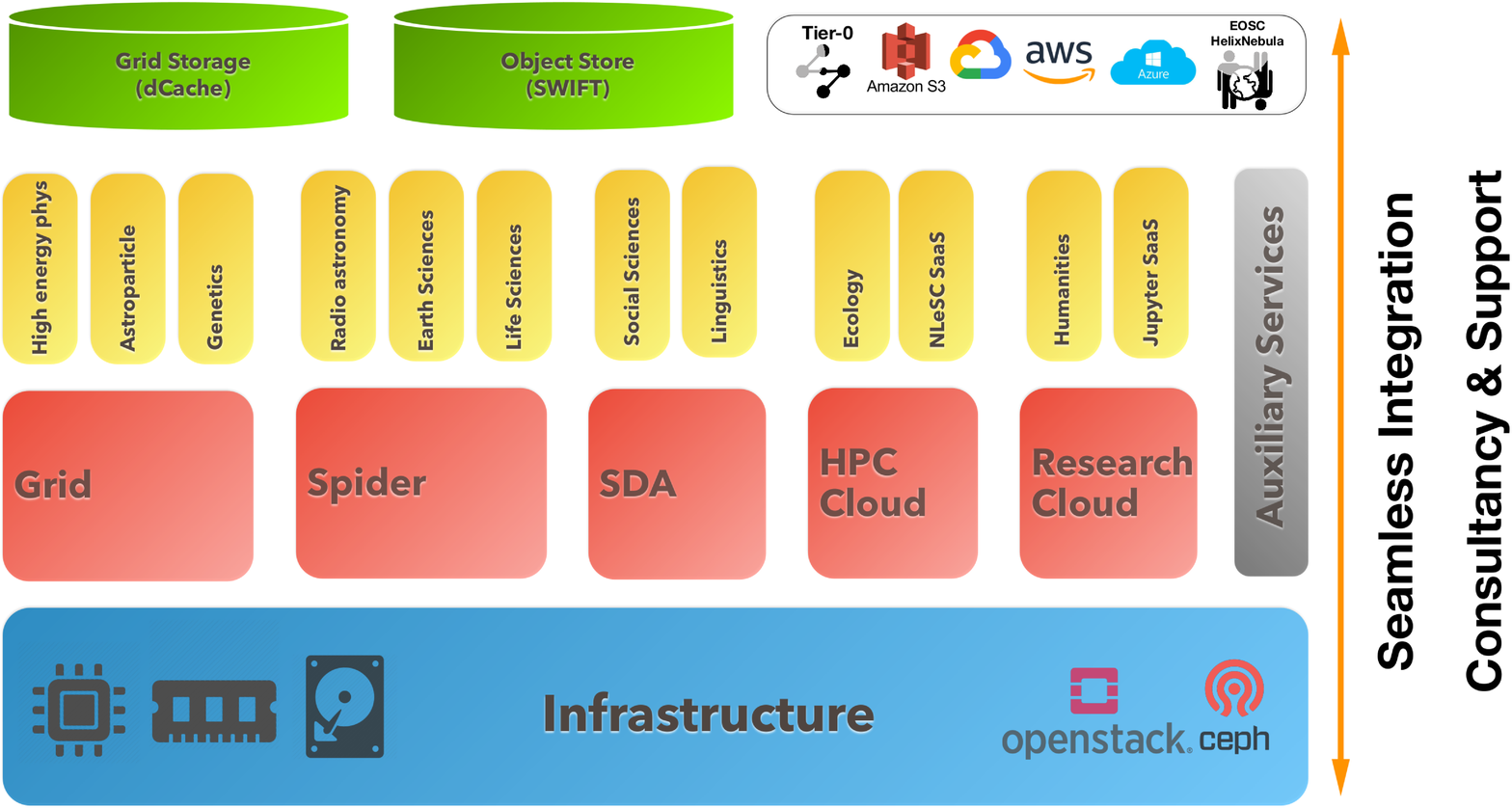}{fig1}{The SURF Virtual Data Center for Data Processing. In the VDC we pool together all compute nodes, virtualise them and provision them via cloud technology (blue layer). On top of these resources, compute platforms (red layer) can easily be deployed and used by different user projects \& communities (yellow: some examples are given). Both internal (blue layer: Ceph) and external (green) storage systems are available.}

The managed cloud approach to the VDC enables us to dynamically scale resources between platforms, thus delivering both efficiency and functionality for researchers. The network and storage systems are considered integral parts of the VDC. Each compute node has access to fast, local storage (e.g., NVMe SSDs). The nodes are connected internally by a high speed (1.2~Tbit/s) EVPN network. This network also connects the nodes to the large internal (Ceph) and external (e.g., dCache and SWIFT) storage systems, and potentially other services external to the VDC. The storage solutions, in combination with the fast (possibly dedicated) network links to the outside world, ensure that the VDC can provide highly efficient data processing platforms with a high level of interoperability.

\section{The Spider platform}\label{Spider-MS4}\label{}
The Grid data processing platform at SURF has a throughput of 400 Petabytes per year. In Q3 2019 it became the first PaaS at SURF to move to the VDC for full operations. This platform, although extremely powerful for data processing, lacks elements of interactivity and ease of use/standard tools that are considered very valuable by researchers that traditionally have not been using the Grid.

To better support data processing in research, SURF has therefore developed Spider (http://doc.spider.surfsara.nl) on the VDC. Spider is especially suited for collaborative teams, that need to process large data sets and that have, or want to develop, sustainable processing pipelines.

\articlefigure{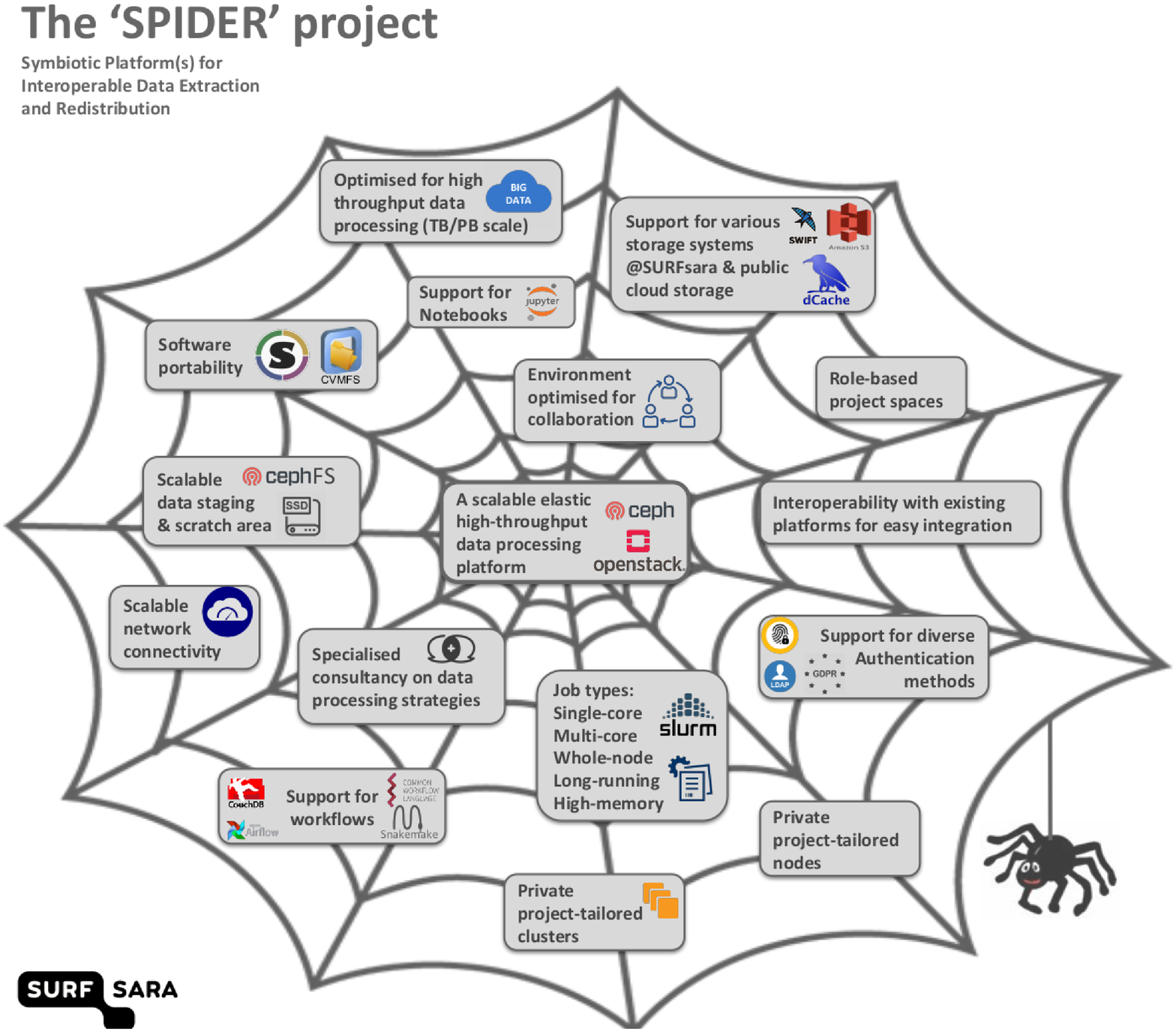}{fig2}{Spider -- a SURF ecosystem for flexible and scalable data processing.}

Fig.~\ref{fig2} shows the current, generic implementation of Spider. The platform is constructed based on individual building blocks that are weaved together using Ansible (https://www.ansible.com). These building blocks are clonable and can be customized. Hence, it is easy to deploy new flavours of Spider for projects with requirements (e.g., in terms of data security) that can not be met on the generic Spider instance. To aid researchers in creating new services on the VDC, e.g. new Spider flavours, SURF offers the MS4 co-creation and service hosting platform. Researchers can propose for access to both Spider and MS4 starting in 2020 (https://www.surf.nl).

\section{Preparing for Exascale Data Processing in Radio Astronomy}\label{proc-astro}
Astronomy, and in particular Radio Astronomy, is one of the largest data generating sciences. Current state-of-the-art radio telescopes, such as LOFAR \citep{vanHaarlem2013}, generate several Petabytes of data products per year and their archives now constitute the worlds largest astronomical data collections. At the same time, new algorithms to calibrate and image this data, imply that radio astronomy is going through a software renaissance. This combination sets important requirements, e.g. in terms interactivity, run-time, disk space, memory and bandwidth, for compute platforms that are considered suitable by individual radio astronomers for their processing workflows.

For LOFAR some advances have been made towards achieving, more robust and automated pipeline platforms \citep[e.g.][]{mechev2018,shimwell2019}. In this setup, small processing teams (continuously) serve science ready data products to the larger community. However, this is not yet a standard practice within the field. The challenges in handling the Petabyte-scale radio data today and the Exabyte-scale that will be generated by the SKA, have been recognised and are investigated in various global and European projects, such as the AENEAS project and the European Open Science Cloud initiatives EOSC-hub and Escape.

We envision that managed PaaS solutions, such as Spider and MS4, running within a scalable VDC environment will provide a first step in achieving a more flexible and scalable high throughput data processing environment for radio astronomy. The Spider platform, in beta mode, has been successfully tested in 2019, with a.o., a variety of LOFAR workflows, and was found to ready to be rolled out further in 2020.

\section{Summary}
At ADASS 2019, SURF presented their latest developments in high performance and high throughput computing. These are particularly relevant for large data generating sciences, such as Astronomy, and we highlight here our new Spider and MS4 platforms.

In addition, SURF also participated in ADASS 2019 with a booth and in a birds of a feather (BoF) session. Here we provided an overview of our services, a demo on long-haul data transfers and the use of hardware accelerators. For more information about SURF and our services please visit surf.nl or contact us via helpdesk@surfsara.nl.

\acknowledgements JBRO would like to thank all SURF staff, especially the Distributed Data Processing group and the Data Processing Services unit at SURFsara.

\bibliography{P6-1_v9}  

\end{document}